\documentclass[twocolumn]{aastex61}
\usepackage{graphicx}
\usepackage{float}
\usepackage{textcomp}
\usepackage{epstopdf}
\usepackage{soul}
\usepackage{xcolor}
\usepackage{hyperref}
\usepackage{lineno}

\received{}
\revised{}
\accepted{}

\submitjournal{MNRAS}

\shorttitle{The Uranus Trojans}
\shortauthors{Wood}
\newcommand{\textedit}{}
\newcommand{\textedittwo}{}

\begin{document}

\title{The Stability of Uranus Trojans Over the Age of the Solar System}

\author[0000-0003-1584-302X]{Jeremy Wood}
\affiliation{Department of Space Studies, American Military University, 111 W. Congress Street, Charles Town, WV 25414}

\correspondingauthor{Jeremy Wood}

\begin{abstract}
\textedit{The stability of eight nominal fictitious Uranus Trojan orbits over the age of the Solar system has been measured.} The initial inclinations, $i_0$, were 0$^{\circ}$, 5$^{\circ}$, 15$^{\circ}$, and 30$^{\circ}$ relative to the ecliptic plane. \textedit{Initial eccentricities ranged from 0 to 0.1 for $i_0= 0^{\circ}, 5^{\circ}$, and 0 to 0.2 for $i_0= 15^{\circ}, 30^{\circ}$.} Half of the orbits were in the L$_4$ swarm, and half were in the L$_5$ swarm. \textedit{Orbits in the L$_4$ swarm had mean longitudes 8.8$^{\circ}$ from the nominal L$_4$ Lagrange point, and orbits in the L$_5$ swarm had mean longitudes 18.2$^{\circ}$ from the nominal L$_5$ point.} I integrated 10,000 massless clones per nominal orbit in the six-body problem (Sun, test particle, four giant planets) for 4.5 Gyr and calculated the half-life for each orbit.
A total of 1291 test particles survived for the entire integration time. Of these survivors, 99\% were associated with the nominal orbit with $i_0=0^{\circ}$ in the L$_4$ swarm. These surviving test particles had initial eccentricities in the range $e_0<0.07$. The half-lives associated with L$_4$ orbits were 1258 Myr, 286 Myr, 56 Myr, and 237 Myr for nominal orbits with $i_0=0^{\circ},5^{\circ},15^{\circ}$, and $30^{\circ}$, respectively. The half-lives associated with L$_5$ orbits were 103 Myr, 281 Myr, 25 Myr, and 46 Myr, respectively. The overall results showed that the ecliptic plane is one good place to search for primordial Uranus Trojans. 
\end{abstract}

\keywords{minor planets, asteroids} %taken from the MNRAS list of keywords

%\linenumbers
%\onecolumn
\section{Introduction}
Trojan asteroids are one example of a reservoir of small solar system bodies (or SSSBs). Trojans librate about the L$_4$ (leading) and L$_5$ (trailing) triangular Lagrange points of a planet \citep[e.g.][]{2019dsss.book.....W,2020PASP..132j2001H} \textedit{that were theorized to exist by \citet{Lagrange_J_L_1772}}.\par 
Technically, the L$_4$ and L$_5$ triangular Lagrange points only exist in the circular restricted three-body problem, however, in the real Solar system problem, SSSBs can still librate about both of them \citep[e.g.][]{2002Icar..159..328M,2006MNRAS.367L..20H,2019dsss.book.....W}. The L$_4$ and L$_5$ Lagrange points are located 60$^{\circ}$ ahead and 60$^{\circ}$ behind the planet in its circular orbit around the Sun, respectively. In the actual Solar system, the centers of libration are offset from these exact locations \citep[e.g.][]{2009MNRAS.398.1715L}.\par
As a Trojan librates about the L$_4$ or L$_5$ Lagrange point, it \textedit{is} in what is called a tadpole orbit because over time, an observer on the planet would see the asteroid librate about the Lagrange point within a region bound by an overall shape that resembles a tadpole \citep[e.g.][]{1999ssd..book.....M, 2002aste.book..725M,2006MNRAS.367L..20H,2019dsss.book.....W}.\par
The question of the stability of tadpole orbits in the circular restricted three-body problem was answered by Lagrange who showed that they could be stable. \textedit{\citet{Gascheau...1843} showed that for a system in which the mass of the Trojan is $<$ the mass of the planet \textedittwo{and their combined mass is $\ll$ the mass of the star, then} the system is linearly stable only if the combined mass of the Trojan and planet does not exceed about 3.7\% of the stellar mass. Under certain conditions, this value can be increased to 3.9\% \citep{2010CeMDA.107..145S}}. It would not be until the 20th century until numerical simulations would show that a tadpole orbit can be stable over the age of the Solar system \citep{1993AJ....105.1987H,1997Natur.385...42L,2020MNRAS.495.4085H} and the discovery of the first Trojan asteroid \citep{1906AN....170..353W}.\par
Since that initial discovery, thousands more Trojan asteroids have been found. Today, we know that the planet Jupiter has approximately 12,000 known Trojans\footnote{https://ssd.jpl.nasa.gov/tools/sbdb\_query.html (accessed July 1, 2022)}, Mars has 9, Saturn has zero, Earth has 2, Uranus has 1 \citep{2002Icar..159..328M,2005Icar..175..397S,2011Natur.475..481C, 2011MNRAS.412..537L,2013Sci...341..994A,2014MNRAS.439.2970D} and Neptune has 32\footnote{https://minorplanetcenter.net/iau/lists/Trojans.html (accessed July 1, 2022)}.\par
The properties of Trojans are important to know because Trojans aid in our understanding of the nature and evolution of the early solar system \citep{1998AJ....116.2590G,2004Icar..167..347K,2005Natur.435..459T,2009MNRAS.398.1715L, 2011epsc.conf.1246L, 2014EPSC....9...54M,2015Icar..247..112P,2016A&A...592A.146G,2018NatAs...2..878N}. \par
One way they can do this is through their composition. \textedit{Trojans are rocky bodies with low visual albedos (the average is 0.074 for Jupiter Trojans\footnote{https://ssd.jpl.nasa.gov/tools/sbdb query.html (accessed Oct. 25, 2022)}), strong infrared spectra, and small amounts of organics \citep{1980Natur.283..840G,2017AGUFM.P23A2708M}.} Since Trojans are leftover debris from the formation of the solar system, their composition can yield vital information about the composition of the Solar nebula where they formed and, if the Trojan formed in situ, the interior of their associated planet \citep{2002aste.book..725M,2003Icar..162..453M,2004Icar..167..347K,Martin...personal...}.\par \textedit{Composition of Trojans can also place constraints on theories of Solar system formation. For example, if near infrared spectra of Jupiter Trojans show that they can be categorized into two unique spectral groups then this may indicate that those Trojans originated in two different regions. If true, then any theory of planetary migration would have to account for that \citep{2017AGUFM.P23A2708M}.}\par 
\textedit{In addition to forming in-situ, asteroids could have also been captured into the Trojan region via various mechanisms such as gas drag friction, rapid increase in the mass of the parent planet, collisional diffusion, and chaotic capture during planetary migration. The latter occurs as secondary resonances sweep over the co-orbital region, making the region chaotic and allowing for small bodies to flow out of and into the Trojan region. Any small body then entering the Trojan region may become permanently captured when the secondary resonances exit the region, and planetary migration ends \citep{2002aste.book..725M,2005Natur.435..462M,2009AJ....137.5003N}.} Henceforth, I will refer to Trojans that remained after planetary migration as primordial Trojans.\par
The orbital properties of primordial Trojans today can place constraints on theories related to Solar system formation and in particular, the migration of the giant planets \citep{1998AJ....116.2590G,2004Icar..167..347K,2009MNRAS.398.1715L,2014Icar..243..287D,2016CeMDA.125..451H,2018ASPC..513...57L,2019ESS.....431201M,2019A&A...631A..89P,2019P&SS..169...78Y}.\par \textedit{For example, the current inclination distribution of Jupiter Trojan orbits extends up to 40$^{\circ}$, and this broad range is best explained using chaotic capture \citep{2005Natur.435..462M,2007MNRAS.380..479M}, and the libration amplitude distribution of the Trojan resonant angle is best explained using long-term dynamical diffusion and capture from the primordial disk \citep{1997Natur.385...42L,2003Icar..162..453M}.} Though it should be noted that the number of primordial members of any Trojan reservoir today \textedit{will} be smaller than it was at the end of all planetary migration.\par
This is because the number of SSSBs in any Trojan reservoir slowly decreases over time through the combined influence of collisions, non-gravitational perturbations \citep{1996Icar..119..192M,1997Icar..125...39M}, and gravitational perturbations by the planets \textedit{such as those due to overlapping resonances that cause Trojan orbits to become unstable by pumping up the amplitude of the librating Trojan resonant angle until the SSSB is ejected from the 1:1 mean motion resonance with the planet \citep[e.g.][]{2004Icar..167..347K,2005CeMDA..92...71T,2008Icar..196....1L,2010MNRAS.405...49H,2013CeMDA.117....3E,2016MNRAS.460.1094M}.}\par
 Therefore, over time, members gradually leak out of their relatively stable reservoirs into dynamically unstable orbits \citep{1983Icar...56...51W,1983Metic..18..422W,2005Natur.435..459T,2010MNRAS.402...13H,2014Icar..243..287D}.\par
\textedit{The rates of escape for each Trojan swarm need not be the same. For example, the escape rate for Jupiter's real Trojans is 23\% for L$_4$ members and 28.3\% for L$_5$ members over 4.5 Gyr. This is despite the fact that the stable regions are symmetrical about L$_4$ and L$_5$ \citep{1998AJ....116.2590G,2002Icar..160..271N,2010CeMDA.107...51D,2014Icar..243..287D,2020A&A...633A.153Z}. Though Jupiter's L$_4$ and L$_5$ Trojans display no large statistical differences among absolute magnitudes, eccentricities, lightcurve amplitudes, periods, and albedo, the number of L$_4$ Trojans is about 1.9 times larger than the number of L$_5$ Trojans \citep{2020IAUS..345..345S}.}\par
\begin{figure} [h]
\centering
\includegraphics[width = \columnwidth]{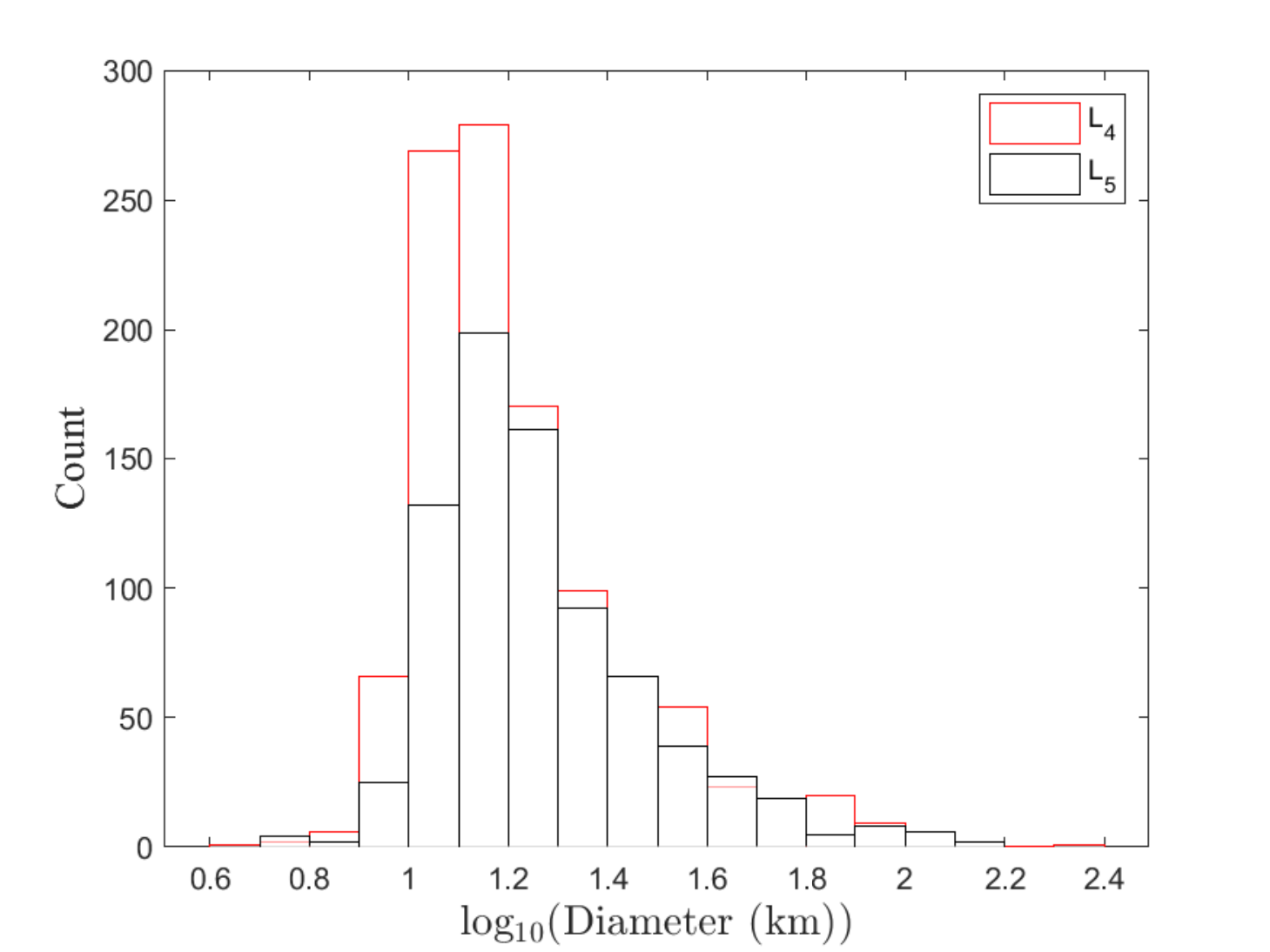}
\caption{\textedit{A histogram of the log$_{10}$(diameters (km)) of the known L$_4$ (red) and L$_5$ (black) Jupiter Trojans.}}
\label{jupiter_trojans_histogram}
\end{figure}

\textedit{The size distribution of the known L$_4$ and L$_5$ Jupiter Trojans is shown in Figure \ref{jupiter_trojans_histogram}.\footnote{https://ssd.jpl.nasa.gov/tools/sbdb\_query.html (accessed July 1, 2022)} The larger number of L$_4$ Trojans compared L$_5$ Trojans for log$_{10}$(diameter (km)) values less than 1.4 (or diameter values less than 25 km) is quite noticeable.}\par

\textedit{This well-known asymmetry between the number of L$_4$ and L$_5$ Jupiter Trojans may simply be primarily primordial and related to how Trojans were captured. Thus, the asymmetry can be explained by differences in initial conditions \citep{2014Icar..243..287D,2020MNRAS.495.4085H}. Another contributing factor to the asymmetry is due to L$_4$ orbits being more stable than L$_5$ orbits during planetary migration \citep{1998AJ....116.2590G}.}\par
\textedit{Collisional history may also be responsible for some of the asymmetry. Jupiter's L$_4$ Trojans have more families  and higher inclinations compared to Jupiter's L$_5$ Trojans. Furthermore, the inclination distribution for L$_4$ Trojans has a noticeable peak at $\approx 8^{\circ}$ with respect to the invariable plane that the inclination distribution of the L$_5$ Trojans does not have \citep{2014Icar..243..287D}. Since asteroids with diameters below 20 km that are members of the Eurybates and Epeios families strongly contribute to this peak, this hints at a collisional explanation as asteroid families are thought to be formed via collisions \citep{2020IAUS..345..345S}.}\par 
\textedittwo{Though Jupiter does have the largest Trojan population of any planet by far, more Trojans of other giant planets may be discovered in the future with their own properties. This is just one reason why it is a good idea to study the properties of real or even fictitious Trojan populations of the other giant planets.}\par
Another factor that must be considered for any Trojan population is that not all Trojans seen today need be primordial. Some of them could also be transient visitors captured into a planet's Trojan region. Thus, a planet's current Trojan population could consist of both primordial and transient members \citep{2000Icar..148..479F,2003A&A...410..725M,2005Icar..175..397S,2012MNRAS.423.2587H,HornerJ:2012a,2013MNRAS.432L..31D,2013DPS....4550802G,2017MNRAS.466..489B,2017NatAs...1E.179P}. \textedittwo{These} transient members \textedittwo{could consist of temporarily captured Centaurs (SSSBs that orbit the Sun among the giant planets) and} would be in orbits that are unstable over the age of the Solar system \citep{2006MNRAS.367L..20H}. \textedittwo{Transient Trojans could even enter the Centaur population, enhancing their number.} In fact, \textedittwo{simulations show that Neptune} Trojans could be a \textedittwo{significant} source for the Centaurs \citep{2010MNRAS.402...13H,2010MNRAS.405...49H}. It is likely that some Centaurs today are former Trojans.\par
\textedittwo{There is also some evidence that there is a relationship between Centaurs and Uranus Trojans.} For example, the Centaur 2014YX$_{49}$ may have been a transient Uranus Trojan in the past \citep{2017MNRAS.467.1561D}. The one known Uranus Trojan, 2011 QF$_{99}$, \textedit{was discovered by \citet{2013Sci...341..994A} while searching for Trans-Neptunian Objects in a 20-square-degree area using the 3.6 meter Canada–France–Hawaii Telescope. 2011 QF$_{99}$} is believed to be a transient, and dynamical simulations suggest that its tenure as a Uranus Trojan will likely be about 1 Myr \citep{2006MNRAS.367L..20H,2013Sci...341..994A,2017MNRAS.467.1561D}. The orbital parameters of this Trojan \textedit{with 1-$\sigma$ errors} are shown in Table~\ref{uranus_trojan}\footnote{https://ssd.jpl.nasa.gov/tools/sbdb\_query.html (accessed 4/20/22)}. \textedit{The symbols $a, e, i, \omega, \Omega$, and $M$ represent the semimajor axis, eccentricity, inclination relative to the ecliptic plane, argument of perihelion, longitude of ascending node and mean anomaly, respectively.} \par
\textedittwo{Like the orbit of this one known Uranus Trojan, it is believed that most Uranus Trojan orbits are unstable over the age of the Solar system \citep{1993AJ....105.1987H,2003A&A...410..725M,2002Icar..160..271N}. This instability is mainly due to overlapping secular and three-body resonances and the near 2:1 commensurability between Neptune and Uranus} \citep{2002Icar..160..271N,2003A&A...410..725M}.\par
\textedittwo{This idea is supported by observations.} \textedit{Other targeted searches have been unable to find any primordial Uranus Trojans.} \textedit{\citet{2003MSAIS...3...40B} surveyed the region offset from each \textedit{Uranian} Lagrange point by $\pm30^{\circ}$ in declination and $\pm$2 hours in right ascension using the 2.2 m telescope of ESO at La Silla. \citet{2006DPS....38.4403S} searched tens of square degrees around the L$_4$ and L$_5$ Lagrangian regions \textedit{of Uranus} using the 6.5 meter Magellan telescope.}\par
\textedit{The Outer Solar System Origins Survey (or OSSOS) searched eight 21-square-degree areas near the invariable plane and at moderate latitudes from the invariable plane using the 3.6 meter Canada–France–Hawaii Telescope. Though their targets were Trans-Neptunian Objects, their search area included regions near Uranus where Trojans could be \citep{2016AJ....152...23V}.\footnote{http://www.ossos-survey.org/about.html (accessed Oct. 19, 2022)} New Trojans \textedittwo{of Uranus or any giant planet} may be discovered by the Decam Ecliptic Exploration Project (DEEP) \citep{2021DPS....5320204T,2022AAS...24022705G}.}\par
Despite the \textedittwo{lack of discoveries}, some small stable niches in phase space have been found which could support primordial Trojans for Uranus \citep{2002ApJ...579..905M,2002Icar..160..271N,2010CeMDA.107...51D,2020A&A...633A.153Z}. \par
Indeed, \citet{2002Icar..160..271N} \textedit{integrated 264 test particles for 4 Gyr with initial eccentricities $e_0<0.1$, initial semimajor axes $19.2\textnormal{ au} \le a_0 \le 19.41\textnormal{ au}$, initial arguments of perihelion $\omega = 240^{\circ} + \omega_U$ (where $\omega_U$ is the argument of perihelion of Uranus), and $0 \le i_{inv} \le 25^{\circ}$ in increments of 5$^{\circ}$, where $i_{inv}$ is the inclination relative to the invariable plane. Values of $\Omega$ were set to that of Uranus, and initial mean longitudes were set to the libration centers for tadpole orbits in the L$_4$ swarm which are $\approx 60^{\circ}$ ahead of that of Uranus for $e_0 = 0$ and higher for $e_0 > 0$ \citep{1999Icar..137..293N}. They found four (or 1.5\%) L$_4$ tadpole orbits with $i_{inv} < 5^{\circ}$ and $e_0=0.01$ that were stable for the entire integration time.}\par 
 \textedit{As another example, \citet{2010CeMDA.107...51D} integrated 160 massless test particles in the 6-body problem (Sun, four giant planets, massless test particle) in the Uranus Trojan region in two separate simulations. In one simulation, initial semimajor axes of test particle orbits were set to 100 values evenly distributed across the range $18.9 \textnormal{ au} \le a \le 19.6\textnormal{ au}$. Initial inclinations were set to $i_{inv}=3^{\circ}$.}\par
\textedit{The argument of perihelion was set to $\omega = \omega_U \pm 60^{\circ}$. Other initial orbital parameters were set to those of Uranus. The integration time was 10 Myr. They found that the center of the stable orbits was located at $a=19.18$ au for L$_4$ Trojans and at $a=19.3$ au for L$_5$ Trojans. The apparent asymmetry was caused by interference from Jupiter.}\par
\textedit{Then a second simulation was performed in which the semimajor axes of test particle orbits were set to $a_U$, the semimajor axis of Uranus, and initial inclinations were set to values in the range $1\le i_{inv}\le60^{\circ}$ in one degree intervals for 60 fictitious Trojans. The integration time was 100 Myr.  In their results they noticed four stability windows: $i_{inv}$ = 0$^\circ$--7$^\circ$, 9$^\circ$--13$^\circ$, 31$^\circ$--36$^\circ$, and 38$^\circ$--50$^\circ$ in which the eccentricity remained at or below 0.2, and no test particles escaped from the Trojan region. Finally, they integrated three fictitious test particles initially placed close to the L$_4$ Lagrange point for 4.5 Gyr with $i_{0}=i_U$, $i_U+2^\circ$, and $i_U+4^\circ$; and all other orbital properties set to those of Uranus. All three were stable over the age of the Solar system.}\par 
Integrations of much larger numbers of Uranus Trojan orbits over the age of the Solar system are rare. \textedit{\citet{2020A&A...633A.153Z}} integrated 35,501 test particles in the Uranus Trojan region for \textedit{4.5 Gyr}. \textedit{Test particles were initially evenly distributed over 131 values of inclination covering the range $0^\circ \le i_{inv}\le 65^\circ$, and 271 values of semimajor axis covering the range 19.08 au -- 19.35 au. Other initial orbital parameters were set to those of Uranus except for $\omega$ which was set to $\omega_U+60^{\circ}$.}\par
\citet{2020A&A...633A.153Z} included planetary migration in their study and reported \textedit{that 3.81\% of the test particles survived the integration time. Of these surviving test particles, 95.5\% had $i_{inv}<7.5^\circ$, and the other 4.5\% had $i_{inv}$ between $42^{\circ}$ and $48^{\circ}$.}\par

\begin{table} [h]
\begin{center}
\begin{tabular}{c c}
\hline
$a$ (au)&19.12$\pm$0.00033\\
$e$&0.1760$\pm$0.0000044\\
$i^{\circ}$&10.80$\pm$0.000044\\
$\omega^{\circ}$&286.79$\pm$0.0042\\
$\Omega^{\circ}$&222.43$\pm$0.00004\\
$M^{\circ}$&306.37$\pm$0.0034\\
\hline
\end{tabular}
\caption{The orbital \textedit{parameters} of the Uranus Trojan 2011 QF$_{99}$ with 1-$\sigma$ errors. Orbital data were taken from the JPL Small Body Database \textedittwo{(accessed Nov. 28, 2022)} for epoch 59800 MJD.}
\label{uranus_trojan}
\end{center}
\end{table}

\textedit{Given the rarity of Giga-year integrations of large numbers of Uranus Trojans,} more integrations of large numbers of Uranus Trojan orbits over the age of the Solar system are needed \textedit{in order to increase our knowledge of the stable regions in the Uranus Trojan region over the age of the solar system}. The purpose of this work is to quantify the orbital stability of Uranus Trojans over the age of the Solar system for four selected initial inclinations and two selected initial mean longitudes.\par
This paper is partitioned as follows. In section 1, I introduce the topic. In section 2, I explain my method. In section 3, I present my results and in section 4 summarize my conclusions.

\section{Method}
\textedit{I used the technique of numerical integration of massless test particles. I started by selecting an initial semimajor axis range for the test particle orbits of 19.2 au--19.4 au.} The value 19.2 au was chosen because I measured it to be the approximate center of libration for the semimajor axis of the orbit of Uranus using planetary orbital data from \citet{2017AJ....153..245W}. The value 19.4 au was chosen because it is the approximate limit of $a$ for tadpole orbits for eccentricities $\le 0.2$ \citep{2002Icar..160..271N}. \textedit{I then created nominal orbits \textedittwo{for eight fictitious Trojans} each with a semimajor axis of 19.3 au. The value of 19.3 au was chosen because it is halfway between 19.2 au and 19.4 au.} \par
Half of the nominal orbits were in the L$_4$ swarm, and half were in the L$_5$ swarm with an initial difference in mean longitude from that of Uranus, $\Delta \lambda_0$, of 51.2$^{\circ}$ and -41.8$^{\circ}$, respectively, \textedit{or orbits in the L$_4$ swarm had mean longitudes 8.8$^{\circ}$ from the nominal L$_4$ Lagrange point, and orbits in the L$_5$ swarm had mean longitudes 18.2$^{\circ}$ from the \textedit{nominal} L$_5$ point.} Mean longitude, $\lambda$, is given by:

\begin{equation}
\lambda = \Omega + \omega + M
\end{equation}

\textedit{An asteroid is considered to be a Uranus trojan if the resonant angle defined by}

\begin{equation}
\phi = \lambda - \lambda_U
\end{equation}

\noindent \textedit{librates in time, where $\lambda_U$ is the mean longitude of Uranus.} In each swarm, there were four nominal orbits with initial inclinations, $i_0$, of 0$^{\circ}$, 5$^{\circ}$, 15$^{\circ}$, and 30$^{\circ}$ relative to the ecliptic plane. $i_0=15^{\circ}$ was deliberately chosen for its known instability \citep{2000DPS....32.1902N,2002Icar..160..271N,2003MNRAS.345.1091M,2010CeMDA.107...51D} (The difference between my $i_0$ and $i_{inv}$ was accounted for \citep{2012A&A...543A.133S}). I wanted to quantify the relative stability between an orbit known to be unstable and other orbits. \par
The initial eccentricities were 0.05 for initial inclinations of 0$^{\circ}$ and 5$^{\circ}$; and 0.1 for initial inclinations of 15$^{\circ}$ and 30$^{\circ}$. Each nominal orbit had an initial longitude of ascending node, $\Omega_0$, and initial argument of perihelion, $\omega_0$, of zero.\par
The $\Delta \lambda_0$ values for orbits in each swarm were different so that the effect of $\Delta \lambda_0$ on stability could be explored. I can do this since the phase spaces around L$_4$ and L$_5$ are dynamically symmetrical to each other \citep{2002Icar..160..271N}. So\textedit{,} in each pair of nominal orbits with the same $i_0$, the only difference between the two orbits was $\Delta \lambda_0$\textedit{, caused only by a difference in mean anomaly}.\par
\begin{figure} [h]
\centering
\includegraphics[width = \columnwidth]{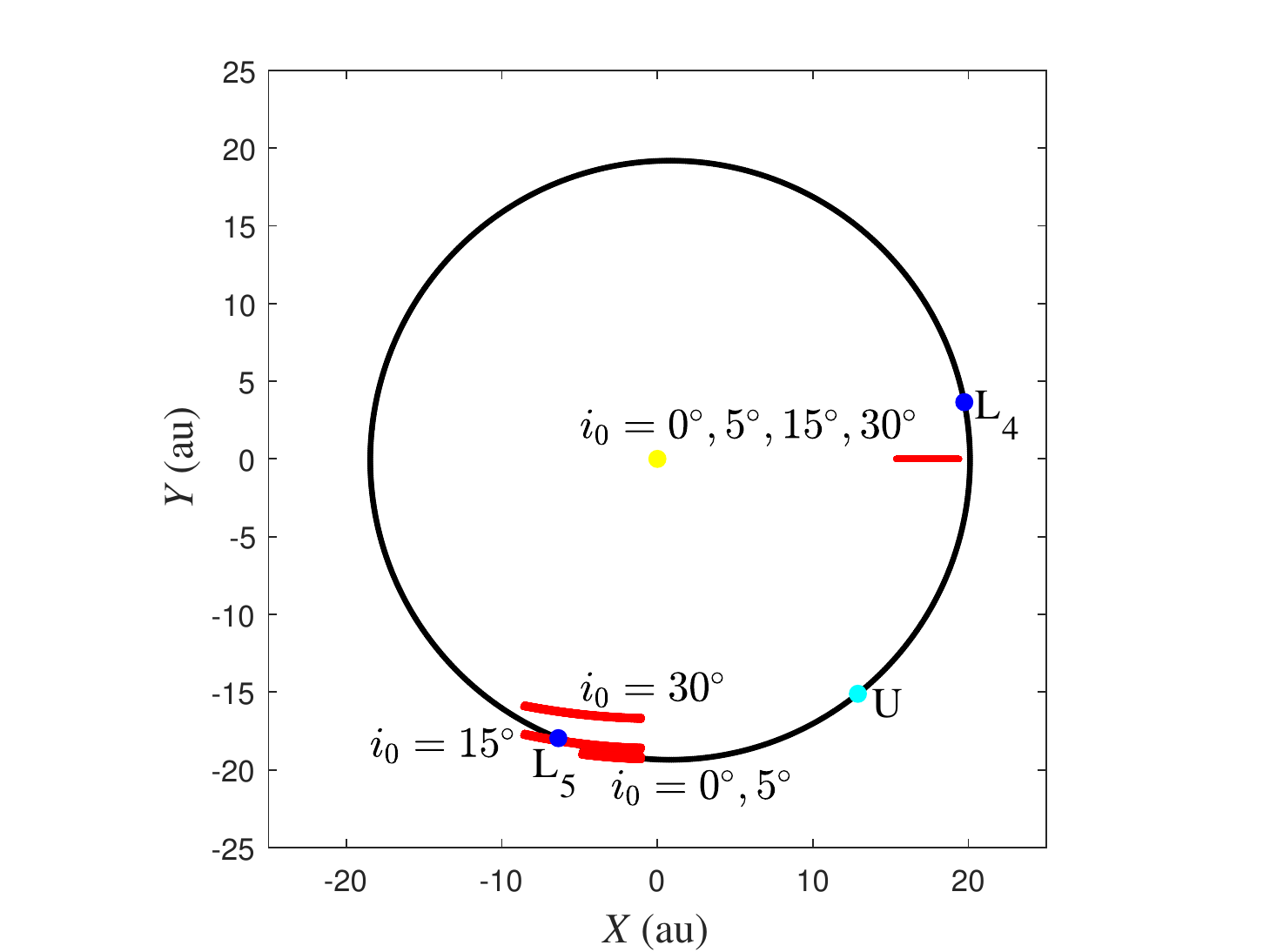}
\caption{\textedit{The initial positions of the test particles used in my simulations in two dimensions ($X-Y$ plane) are shown in red and labeled by inclination. Also shown are:  the planet Uranus, labeled U, its heliocentric orbit projected onto the $X-Y$ plane, the Sun as the yellow circle at the origin, and the nominal positions of the L$_4$ and L$_5$ Lagrange points.}}
\label{tp_initial_xyz}
\end{figure}

  
\textedit{Differences in initial conditions like this can cause differences in the libration amplitude of each test particle about the Lagrange point. Niches of stability can be found in libration-amplitude space \citep[e.g.][]{2009MNRAS.398.1715L,2010CeMDA.107...51D}. The initial positions of the test particles are shown in Figure~\ref{tp_initial_xyz}}.\par

\textedit{The $\Delta \lambda_0$ values are random, so} there is nothing particularly special about \textedit{them} except that they place the test particles within \textedit{20}$^\circ$ of the L$_4$ or L$_5$ Lagrange point which is well within the  \textedit{60-degree angular distance between a planet and the nominal location of a Lagrange point and} so are appropriate for study. Note that no stable horseshoe orbits exist for Uranus Trojans over the age of the Solar system \citep{2020A&A...633A.153Z}. The initial orbital parameters of the eight nominal orbits are shown in Table \ref{nominal}. \par 

\begin{table} [h]
\begin{center}
\begin{tabular}{ c c c c c }
\hline
Swarm&$a_0$ (au)&$e_0$&$i_0$ (deg)&$\Delta \lambda_0$ (deg)\\
\hline
L$_4$&19.3&0.05&0&51.2\\
L$_4$&19.3&0.05&5&51.2\\
L$_4$&19.3&0.1&15&51.2\\
L$_4$&19.3&0.1&30&51.2\\
L$_5$&19.3&0.05&0&-41.8\\
L$_5$&19.3&0.05&5&-41.8\\
L$_5$&19.3&0.1&15&-41.8\\
L$_5$&19.3&0.1&30&-41.8\\
\hline
\end{tabular}
\caption{The orbital parameters of the eight nominal orbits. $a_0$, $e_0$, $i_0$, and $\Delta \lambda_0$ represent the initial semimajor axis, initial eccentricity, initial inclination with respect to the ecliptic plane, and initial difference in mean longitude from that of Uranus, respectively. The initial longitude of ascending node and argument of perihelion were set to zero for each orbit.}
\end{center}
\label{nominal}
\end{table}

Each orbit was cloned by creating a cluster of test particle orbits evenly distributed about the semimajor axis and eccentricity of each nominal orbit in phase space. The initial eccentricities for each cluster were chosen from 100 values equally spaced apart over the range 0--0.1 for orbits with inclinations of 0$^{\circ}$ and 5$^{\circ}$; and 0--0.2 for orbits with inclinations of 15$^{\circ}$ and 30$^{\circ}$. These eccentricity ranges were based on the work of \citet{2010CeMDA.107...51D}. \textedit{The initial semimajor axes, $a_0$, were chosen from 100 values equally spaced apart over the range 19.2 au--19.4 au.}\par
\textedit{The 100 initial eccentricity values along with the 100 initial semimajor axis values} yielded a total of 100 $\times$ 100 = 10,000 clones for each nominal orbit. A nominal orbit can be understood as being an orbit which has the average values of semimajor axis and eccentricity for a cluster of 10,000 clones. \par
Eight nominal orbits with 10,000 clones each yielded a total of 10,000 $\times$ 8 = 80,000 massless test particles. To my knowledge, this is the largest number of test particles per orbit ever integrated for this long a duration for this part of the Uranus Trojan region. \textedit{The advantage of a larger number of test particles per orbit is increased accuracy in half-life measurement and greater resolution of $a-e$ maps.}\par 
I then integrated the test particles in the six-body problem (Sun, test particle, four giant planets) for 4.5 Gyr using the symplectic Hybrid integrator within the \textsc{Mercury} dynamics package \citep{ChambersJE:1999}.\par
The fractional energy change due to the integrator did not exceed order $10^{-7}$. The error in energy followed \textit{Brouwer\textsc{\char13}s Law} \citep{1937AJ.....46..149B} which states that if the error is due only to random error in calculated values using a limited number of decimal or bit places then the error in energy grows as $\sqrt{t_{int}}$, where $t_{int}$ is the total integration time. The position errors of the planets and small bodies grew as $t_{int}^{\frac{3}{2}}$ in accordance with Brouwer\textsc{\char13}s Law, though I was unable to measure these errors due to time constraints. The fractional angular momentum change due to the integrator did not exceed order $10^{-11}$.\par
Close encounters were accounted for by a switch to an accurate Bulirsch-St$\ddot{o}$er algorithm for close encounters between a test particle and a planet occurring within a distance of three planetary Hill Radii \citep[e.g.][]{ChambersJE:1999,2020AJ....159..179H}. The step time was 40 days or about one-thousandth the orbital period of Uranus, following \citet{2002Icar..160..271N}, and is one-third the time step used by \citet{2004MNRAS.354..798H} for integrations of Centaurs. The output time was set to 100,000 years.\par
Initial orbital parameters of the four giant planets were taken from the NASA JPL HORIZON ephemeris for epoch July 6, 1998 at UT 00:00\footnote{https://ssd.jpl.nasa.gov/horizons/app.html\#/ (accessed 31st December 2015)}. Inclinations were relative to the ecliptic plane. Test particles and planets were fully interacting and subject only to the gravitational forces of the Sun and giant planets.\par
In this study, test particles were removed from the simulation upon colliding with a planet, colliding with the Sun (obtaining a heliocentric distance $\le$ 0.005 au), or obtaining a a heliocentric distance $>$ 100 au. The removal time of each test particle was recorded.\par
\textedit{After a relatively brief period of time in which the swarm disperses,} the number of test particles in a Trojan reservoir decreases roughly exponentially over time \citep[e.g.][]{2004MNRAS.354..798H,2010MNRAS.405...49H}, then the half-life, which is a measurement of stability, can be found by using the standard radioactive decay equation. Given $N_o$ as the initial number of test particles at a time $t = 0$, this equation is:

\begin{equation}
N = N_0e^{-\rho t}
\label{half_life_eqn}
\end{equation}

\noindent where $N$ is the number of test particles remaining at a time, $t$, and $\rho$ is called the decay constant. The half-life, $\tau$, is related to $\rho$ via the equation: 

\begin{equation}
\tau=\frac{-0.693}{\rho}
\label{lambda_eqn}
\end{equation}

To determine the number of remaining test particles as a function of time, a regular time interval at which the number of remaining test particles was to be sampled was determined. To find this sample time, $t_s$, I used the built-in algorithm of MATLAB \citep{2018MATLAB...R2018a} used to find the best bin size for a histogram made using a set of numbers.\par
This algorithm uses Scott's Rule \citep{Scott...2010Wiley} to find a crude value of the bin size, $t_b$, and then refines it. Scott's Rule is given by:

\begin{equation}
t_b = 3.5\frac{\sigma}{N_0^{\frac{1}{3}}}
\label{sample_time_eqn}
\end{equation}

\noindent where $\sigma$ is the standard deviation of the data set. The algorithm also returns the number of bins, $n_{b}$. The algorithm used to refine the bin size and find $n_{b}$ can be found in the Appendix. \textedit{This technique of using a sample time \textedittwo{can be} an improvement over the \textedittwo{current} method in which no constant sampling time is used. \textedittwo{With the current method, the number of remaining test particles is sampled any time the number of test particles changes, even while the Trojan swarm is dispersing. When the Trojan swarm is still dispersing, the remaining number of test particles does not follow the radioactive decay equation. With my new method, if the sampling time is large enough, some data points that are taken while the swarm disperses can be skipped. This leads to a more accurate determination of the half-life.}}\par
My goal was to find the sample times for each nominal orbit. To accomplish this, I input the set of removal times for test particles starting in each nominal orbit separately into the algorithm to obtain the sample times. A set of $N$(t) data was created for each nominal orbit.\par
The half-life for each set of $N$(t) data was found using the method of \citet{2004MNRAS.354..798H} \textedit{but now with a constant sampling time}. Using this method, the $\chi ^2$ function given by 

\begin{equation}
\chi ^2(\rho) = \sum_{i=1}^{n_b+1}\Bigg{[}\frac{\textnormal{ln}N_i-\textnormal{ln}N_0+\rho t_i}{\sigma_i}\Bigg{]}^2
\label{chi_squared_eqn}
\end{equation}

\noindent is minimized. Here, $N_i$ is the number of test particles remaining at a time, $t_i$, and $\sigma_i$ are the individual standard deviations of the data points. Consecutive values of $t_i$ are separated by the sample time so that $t_{i+1}-t_i=t_s$. The $\sigma_i$ are unknown but are set to a constant value \textedit{(of one)} following \citet{2004MNRAS.354..798H}. To determine the half-life for a certain set of $N$(t) data, first, data points with duplicate values of $N_i$ are removed from the data. Next, $\chi ^2$ is determined using an initial guess for $\rho$. Then, the value of $\rho$ is adjusted until the value of $\chi ^2$ is minimized. Finally, the half-life is found using the final value of $\rho$ in equation \ref{lambda_eqn} that yields the half-life to within 1 Myr. Because of the large width in $a_0$ and $e_0$ used to make the clones, a measured half-life would not be a property of a nominal orbit per se but would rather be a value related to a region of the Trojan cloud in phase space.\par
\textedit{After this, I performed a separate integration to measure the initial full libration amplitude (crest to trough) of the resonant angle, $\phi$, for each test particle. I used the same initial conditions and step time as in the previous integrations, however, I changed the integrator to the IAS15 integrator in the \textsc{REBOUND} N-body simulation package \citep{2012A&A...537A.128R,2015MNRAS.446.1424R}. This non-symplectic integrator has a high accuracy and adaptive timestepping.}\par
\textedit{As before, only gravitational forces were considered. Massless test particles were removed upon collision with a planet or the Sun; or obtaining a semimajor axis outside of the range 18.5 au--21  au. The integration time was 100,000 years. I manually analyzed the resonant angle of 100 test particles over time and determined that this would be more than enough time for one cycle to occur. The masses of the Sun and planets were taken from DDE 440 \citep{2021AJ....161..105P}. Planetary and solar radii were obtained from NASA. The masses of the inner planets were added to the Sun. The output time was 1 year.}

\section{Results}
A total of 1291 test particles (1.61\%) survived for the entire integration time. \textedit{This value is very similar to the value of 1.5\% reported by \citet{2002Icar..160..271N}}. Table~\ref{reason_table} shows the number and percentage of test particles removed from the simulation by reason for removal.

\begin{table} [h]
\begin{center}
\begin{tabular}{ c c c }
\hline
Reason&Total&Percentage\\
\hline
Jupiter&462&0.58\\
Saturn&211&0.26\\
Uranus&75&0.09\\
Neptune&50&0.06\\
Ejection&77911&97.39\\
\hline
\end{tabular}
\caption{The number and percentage of test particles removed from the simulation due to a collision with a planet or ejection due to obtaining a heliocentric distance beyond 100 au.}
\end{center}
\label{reason_table}
\end{table}

\begin{figure} [h]
\centering
\includegraphics[width = \columnwidth]{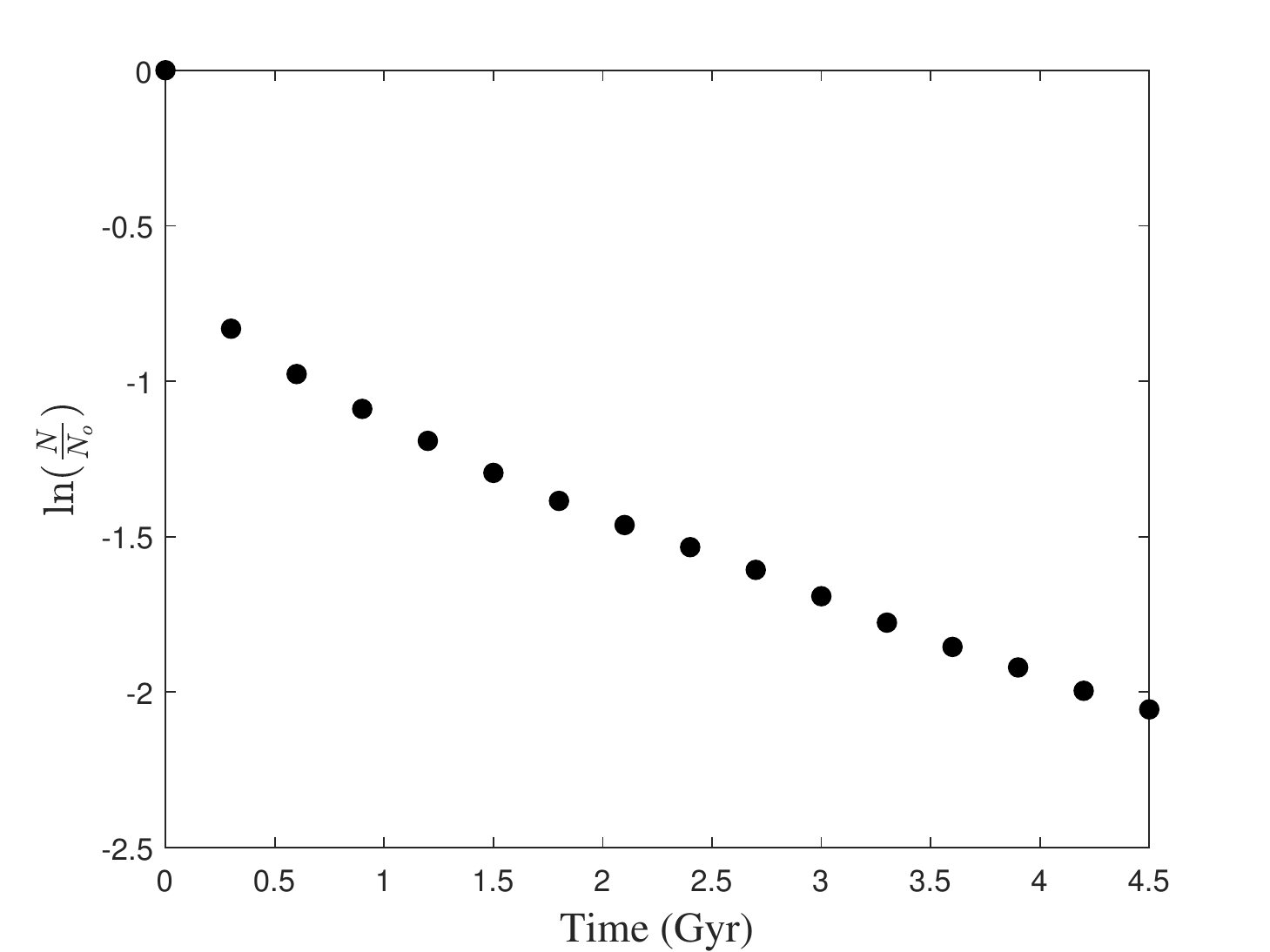}
\caption{A plot of the natural log of the fraction of remaining test particles vs. time for the nominal orbit in the L$_4$ swarm with an initial inclination of $0^\circ$. The sample time was 0.3 Gyr, and the half-life was 1258 Myr.}
\label{half_life_figure}
\end{figure}

\textedit{As expected, the radioactive decay equation well-described the removal of test particles from the simulation for each nominal orbit. }Figure~\ref{half_life_figure} shows a plot of the natural log of the fraction of remaining test particles vs. time for the nominal orbit in the L$_4$ swarm with an initial inclination of $0^\circ$. The non-linearity of this plot at times below 0.5 Gyr is due to the time it takes the test particle orbits to diverge. This was also observed by \citet{2017AJ....153..245W} and \cite{2018AJ....155....2W}. The linear regression coefficient for times $>$ 0.5 Gyr was a very strong -0.998. This strong linearity was also observed for all other nominal orbits with linear regression coefficients ranging from -0.998 to -0.929. The sample time was 0.3 Gyr, and the half-life was 1258 Myr. 

\textedit{Table \ref{table_L4} shows the number of survivors and half-lives for nominal orbits in the L$_4$ and L$_5$ swarms by initial inclination. Of the test particles that survived the integration, 99\% started with $i_0=0^{\circ}$ in the L$_4$ swarm.}\par
\textedit{The survivability for nominal orbits with $i_0>0$ in the L$_4$ swarm was extremely low compared to that of the nominal orbit with $i_0=0^{\circ}$. Though orbits in the L$_4$ swarm with $i_0=0^{\circ}$ had a 12.8\% survivability, orbits in the L$_4$ swarm with $i_0=5^{\circ}$ and $i_0=30^{\circ}$ had only a 0.05\% and 0.02\% survivability, respectively. Not surprisingly, none of the test particles with $i_0=15^{\circ}$ survived the integration.}\par
\textedit{This result is reminiscent of the results of \citet{2002Icar..160..271N} who found that orbits with $i_{inv}=0^{\circ}$ had the highest probability of survival and} \textedit{noticed a similar lopsided difference in survivability compared to orbits with $i_{inv}>0$. They found that orbits with $i_{inv}= 0^{\circ}$ had a 9.1\% survivability, but orbits with $i_{inv}$ in the range $5^{\circ} \le i_{inv} \le  25^{\circ}$ in steps of $5^{\circ}$ had a 0\% survivability over the age of the solar system.}\par 
\textedit{My results} show that the ecliptic plane is one good place to search for primordial Uranus Trojans. \textedit{The lack of discoveries of primordial Uranus Trojans may indicate that if they exist, they are too dark or too small to be detected by current surveys. I suggest that the recently launched James Webb Space Telescope could be used in a targeted search.}\par
\textedit{The difference among half-lives for nominal orbits in the L$_4$ swarm is striking. The ratio of the half-life \textedit{of the nominal orbit with $i_0=0^{\circ}$ to that of the nominal orbits with $i_0=5^{\circ}$,  $i_0=15^{\circ}$, and  $i_0=30^{\circ}$} was approximately four to one, twenty-two to one and five to one, respectively.}\par
For the L$_5$ swarm, there were only four survivors - two with $i_0=0^{\circ}$ and two with $i_0=5^{\circ}$. The nominal orbit with $i_0=5^{\circ}$ had the longest half-life of the swarm of 218 Myr. \textedit{This seems surprising since in the L$_4$ swarm, test particles with $i_0=0^{\circ}$ had the longest half-life. The ratio of this half-life to that of nominal orbits with $i_0=0^{\circ}$, $i_0=15^{\circ}$, and $i_0=30^{\circ}$ was about two to one, eight to one, and four to one, respectively. As expected, the nominal orbit with $i_0=15^{\circ}$ had the shortest half-life just as it did in the L$_4$ swarm.} \par
\textedit{Each nominal L$_5$ orbit had a smaller half-life than the L$_4$ nominal orbit with the same initial inclination. This is likely due to the difference in mean initial libration amplitude of the Trojan resonant angle, $\overline{A}_0$, between the two swarms as shown in Table \ref{table_L4}. The L$_5$ swarm had a mean initial libration amplitude of about 57$^{\circ}$, and the L$_4$ swarm had a mean of about 38$^{\circ}$. For a given eccentricity, on a large enough scale, test particles in orbits with a higher libration amplitude are ejected from the Trojan region at a noticeably faster rate than test particles in orbits with a lower libration amplitude due to chaotic diffusion \citep{2010MNRAS.405...49H,2014Icar..243..287D}.}\par 
\textedit{In this case, I believe that the higher instability of the L$_5$ swarm is due to greater chaotic diffusion caused by the near 2:1 commensurability between the orbital periods of Uranus and Neptune \citep{2002Icar..160..271N}. The current ratio of these orbital periods is 1.96. The greater chaotic diffusion is ultimately due to different initial conditions. Therefore, it was different initial conditions that caused the asymmetry between the two swarms, as it also largely caused the asymmetry observed between Jupiter's Trojan swarms today.} \par
\textedit{Differences in chaotic diffusion can also be used to explain differences in survivability and half-lives among nominal orbits in the same swarm. Interestingly, each nominal orbit that had surviving test particles also had an initial inclination found in an inclination stability window given by \citet{2010CeMDA.107...51D} once $i_0$ was converted to $i_{inv}$.}\par
\textedit{The range of initial eccentricities of surviving test particles are shown in Table \ref{table_L4} and are represented by $e_{0range}$. What is unexpected is that orbits with  $i_0=30^{\circ}$ in the L$_4$ swarm can have stable orbits with $e_0>0.09$. Values of $e_{0range}$ for $i_0=0^{\circ}$ and $i_0=5^{\circ}$ in both swarms are either 0.07 or 0.08. Caution should be used when interpreting these results because $e_{0range}$ values for orbits with $i_0 \ge 5^{\circ}$ are based on a total of only eleven surviving test particles.}

\begin{table} [h]
\begin{center}
\begin{tabular}{c c c c c c c}
\hline
Swarm&$i_0^{\circ}$&Survivors&$\tau$ (Myr)&$ \frac{\tau L_4}{\tau L_5}$&$\overline{A}_0$ (deg)&$e_{0range}$\\
\hline
L$_4$&0&1280&1258&12.2&37.3&$<0.07$\\
L$_4$&5&5&286&1.3&37.2&$<0.08$\\
L$_4$&15&0&56&2.2&38.8&-\\
L$_4$&30&2&237&5.2&39.2&$<0.1$\\
L$_5$&0&2&103&-&58.3&$<0.08$\\
L$_5$&5&2&218&-&58.2&$<0.07$\\
L$_5$&15&0&25&-&59.1&-\\
L$_5$&30&0&46&-&52.5&-\\
\hline
\end{tabular}
\caption{The number of survivors and half-lives, $\tau$, for the L$_4$ and L$_5$ swarms by initial inclination, $i_0$. The ratio of the half-life of an L$_4$ orbit to that of an L$_5$ orbit with the same $i_0$ is also shown. \textedit{The mean initial libration amplitude of the Trojan resonant angle, $\overline{A}_0$, is also shown. The last column shows the range of initial eccentricities for surviving test particles.}}
\end{center}
\label{table_L4}
\end{table}

\begin{figure} [h]
\centering
\includegraphics[width = \columnwidth]{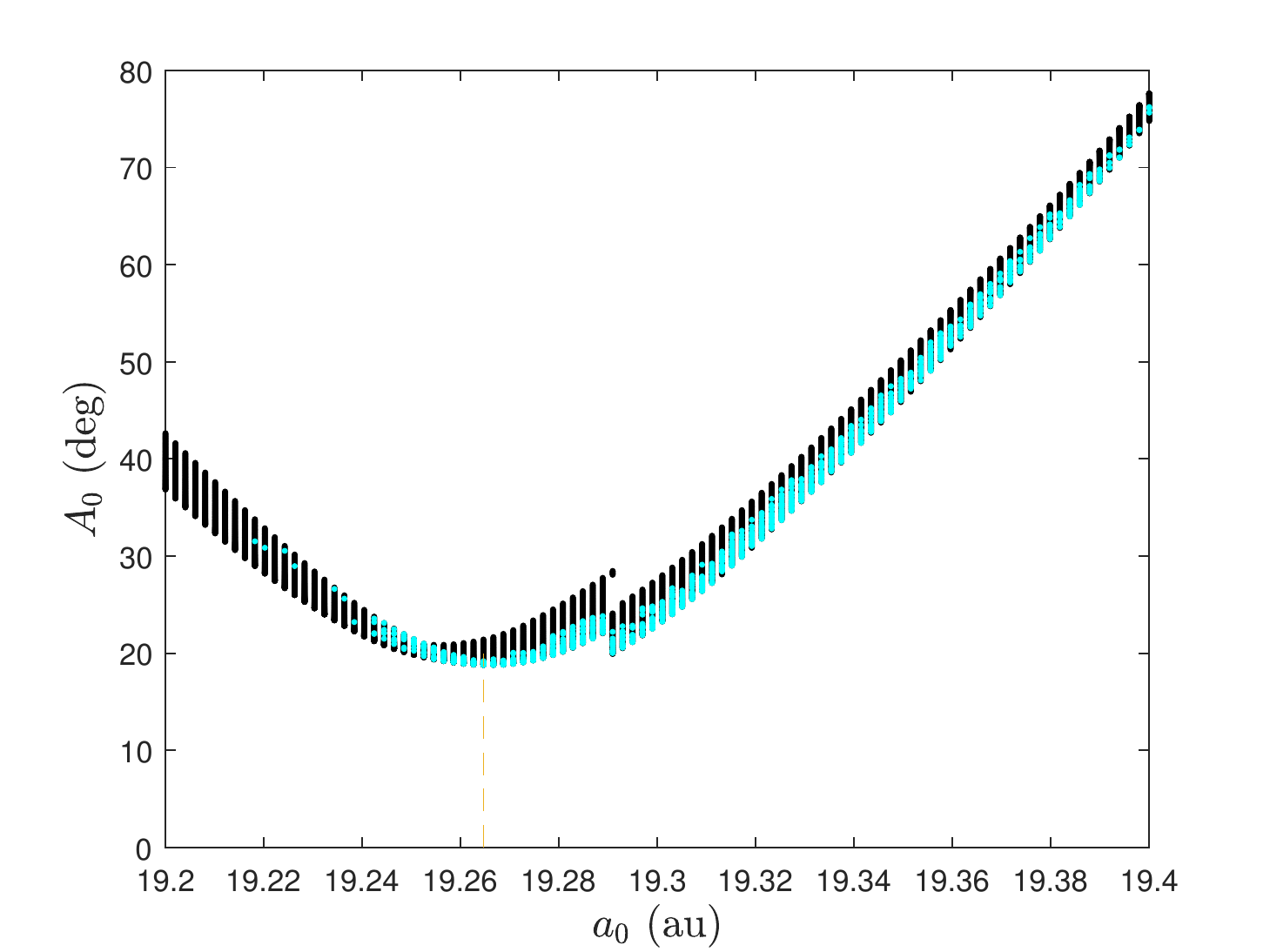}
\caption{\textedit{The initial libration amplitude of the Trojan resonant angle defined by $\lambda - \lambda_U$ vs. initial semimajor axis for test particles in the L$_4$ swarm with $i_0=0^{\circ}$. The vertical dashed line at $a_0=$ 19.265 au is at the semimajor axis associated with the lowest libration amplitude of 18.795$^{\circ}$. \textedittwo{Data points in cyan represent test particles that survived the entire integration, and those in black represent non-survivors.}}}
\label{libration_amplitudes_L4_0}
\end{figure}

Figure \ref{libration_amplitudes_L4_0} \textedit{shows the initial libration amplitude, $A_i$, of $\phi$ vs. initial semimajor axis for test particles in the L$_4$ swarm with $i_0=0^{\circ}$. Data points in cyan represent test particles that survived the entire integration time. The minimum initial libration amplitude of 18.795$^{\circ}$ occurs at a semimajor axis of 19.265 au. This value varied little for other values of  $i_0$, ranging from 19.265 au to 19.289 au over both swarms. The smallest libration amplitude was $14.5^{\circ}$ for an orbit with  $i_0=30^{\circ}$ in the L$_4$ swarm. Nearly all surviving test particles had an $A_i$ value below $80^{\circ}$ in agreement \citet{2010CeMDA.107...51D} who reported that stable orbits over 10 Myr had $A\le 80^{\circ}$. Only two survivors had $A_0>80^{\circ}$:  one from the L$_4$ swarm with $A_0=80.8^{\circ}$ and $i_0 = 30^{\circ}$; and one from the L$_5$ swarm with $A_0=81.3^{\circ}$ and $i_0 = 0^{\circ}$.}

\begin{figure*} [hpt]
\centering
\includegraphics{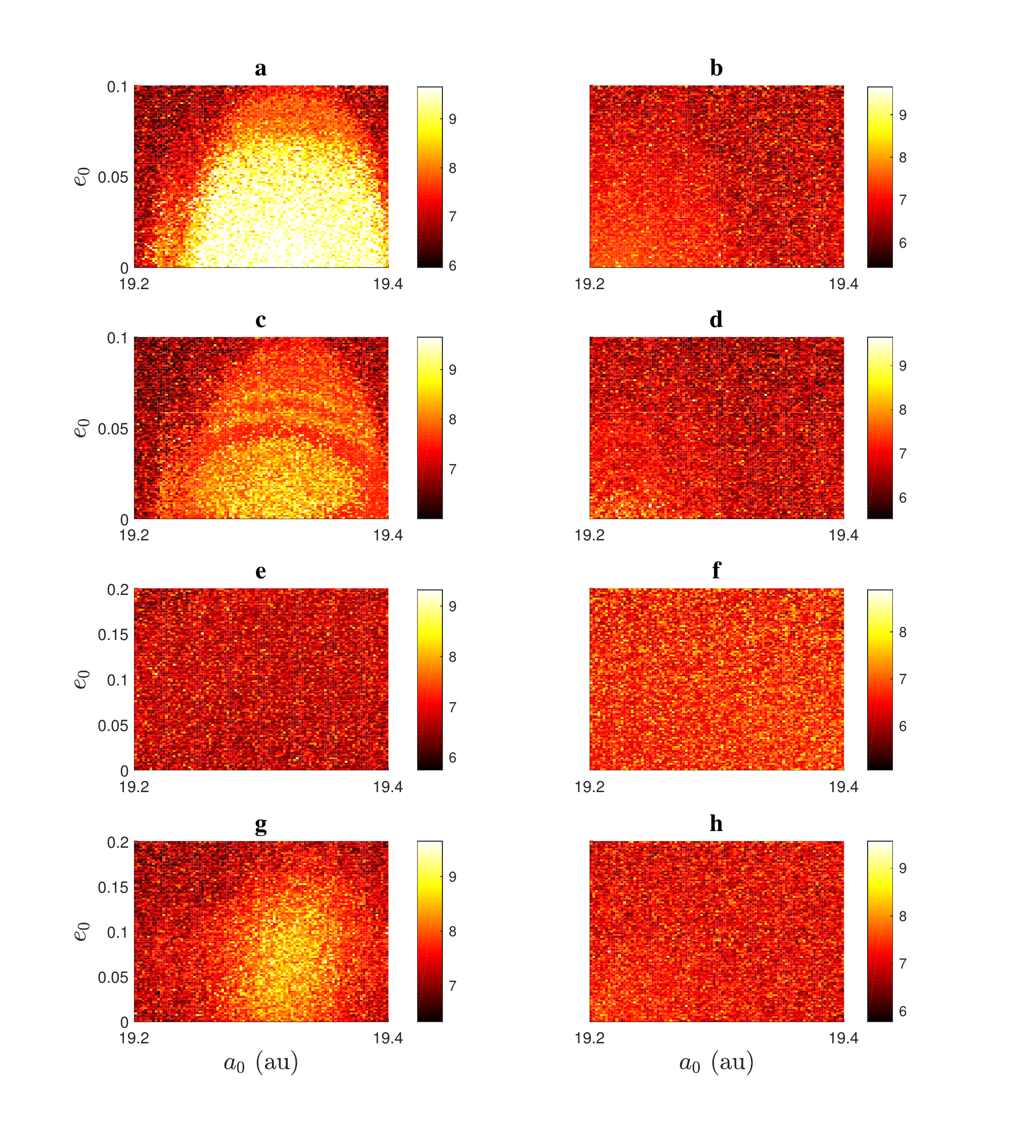}
\caption{Eight panels that each display a map of the dynamical lifetimes in $a_0-e_0$ space. Each map is associated with a nominal orbit. The left column is associated with orbits in the L$_4$ swarm, and the right column is associated with orbits in the L$_5$ swarm. From top to bottom, each row represents nominal orbits with an initial inclination of 0$^{\circ}$, 5$^{\circ}$, 15$^{\circ}$, and 30$^{\circ}$ with respect to the ecliptic plane, respectively. \textedit{Each colorbar has units of} log$_{10}$ of the dynamical lifetime in years.}
\label{dynamical_lifetime_maps}
\end{figure*}

Figure~\ref{dynamical_lifetime_maps} shows \textedit{a map of the dynamical lifetimes in $a_0-e_0$ space for each nominal orbit. L$_4$ orbits are in the left column (panels (a), (c), (e), and (g)), and L$_5$ orbits are in the right column (panels (b), (d), (f), and (h)). Rows from top to bottom correspond to $i_0$ values of $0^{\circ}$, $5^{\circ}$, $15^{\circ}$, and $30^{\circ}$, respectively. Each map was made using 10,000 test particles. As far as I know, these maps have the highest resolution of any $a-e$ map ever made of the Uranus Trojan region over 4.5 Gyr.} Over all panels, the most striking feature is the mound shape in panel (a) that encloses most of the overall surviving test particles. 5372 lifetimes in this panel are above 10$^8$ yr. Surviving test particles in this panel have initial \textedit{semimajor axes in the range $19.2182 \textnormal{ au} \le a_0\le 19.4000\textnormal{ au}$. \citet{2002Icar..160..271N} observed a similar mound shape in their map for a 39.4 Myr integration for initial semimajor axis values between 19.2 au and 19.4 au that extended up to $e_0 \approx  0.15$.}\par 
\textedit{This range differs from the stability region reported by \citet{2020A&A...633A.153Z} of $19.164 \textnormal{ au} \le a_0\le 19.288\textnormal{ au}$ who studied only test particles starting with $e_0$ set equal to the eccentricity of the orbit of Uranus, $e_U=0.04975$. To compare my results to theirs, I examined those surviving test particles in Figure~\ref{dynamical_lifetime_maps} panel (a) with initial eccentricities within 0.001 of that of the orbit of Uranus (none of them had $e_0 = e_U$). I found that their initial semimajor axes were in the range $19.2586 \textnormal{ au} \le a_0\le 19.3798\textnormal{ au}$.}\par
\textedit{My result shows that there can be surviving test particles outside of the range stated by \citet{2020A&A...633A.153Z} for different initial conditions with a small difference in initial eccentricity from $e_U$.}\par
In panel (c), the \textedit{mound-shaped feature that was seen in panel (a)} has been greatly diminished and is contaminated with lifetimes below $10^8$ years. Two mostly horizontal bands of lifetimes below $10^8$ years cut across the feature. These may be due to the presence of secondary and secular resonances as reported by \citet{2020A&A...633A.153Z}.\par
In panel (e), the feature has disappeared completely, and practically all lifetimes are below $10^8$ years. The feature partially returns in panel (g), but it is less prominent with one twenty-fifth the number of orbits with lifetimes $\ge 10^9$ years than in panel (a). Panels (b), (d), (f), and (h) are mostly featureless.\par

\newpage
\section{Conclusions}
The stability of \textedit{eight nominal Uranus Trojan orbits created using} four different initial inclinations \textedit{and two selected initial mean longitudes over the age of the Solar system has been measured.} I integrated 10,000 massless clones per nominal orbit in the six-body problem (Sun, test particle, four giant planets) for 4.5 Gyr and measured the stability associated with each orbit by calculating the half-life.\par
Half of the nominal orbits were in the L$_4$ swarm, and half were in the L$_5$ swarm with an initial difference in mean longitude from that of Uranus, $\Delta \lambda_0$, of 51.2$^{\circ}$ and -41.8$^{\circ}$, respectively. \textedit{Orbits in the L$_4$ swarm had mean longitudes 8.8$^{\circ}$ from the nominal L$_4$ Lagrange point, and orbits in the L$_5$ swarm had mean longitudes 18.2$^{\circ}$ from the nominal L$_5$ point.} In each swarm, there were four nominal orbits with initial inclinations, $i_0$, of 0$^{\circ}$, 5$^{\circ}$, 15$^{\circ}$, and 30$^{\circ}$ relative to the ecliptic plane.\par
A total of 1291 test particles (1.61\%) survived for the entire integration time, and 1280 (99\%) of these survivors started in the nominal orbit in the L$_4$ swarm with $i_0 = 0^{\circ}$. These surviving test particles had initial eccentricities in the range $e_0<0.07$ \textedit{and initial semimajor axes in the range $19.2182 \textnormal{ au} \le a_0\le 19.4000\textnormal{ au}$}. \textedit{The lack of surviving test particles for other nominal orbits is likely because of the near 2:1 commensurability between the orbital periods of Neptune and Uranus which causes chaotic diffusion}. Orbits with $i_0 =15^{\circ}$ were the most unstable for both swarms. No test particles with $i_0=15^{\circ}$ survived the integration. \textedit{Nearly all ($>99\%$) surviving test particles started with an initial Trojan resonant angle libration amplitude below $80^{\circ}$.}\par
L$_4$ orbits with $i_0 = 0^{\circ}$, $5^{\circ}$, $15^{\circ}$, and $30^{\circ}$ had half-lives of 1258 Myr, 286 Myr, 56 Myr, and 237 Myr, respectively. The fact that nearly all surviving test particles and the longest half-life were associated with the L$_4$ orbit with $i_0 = 0^{\circ}$ shows that the ecliptic plane is one good place to search for primordial Uranus Trojans. \textedit{The fact that no targeted searches have found any primoridal Uranus Trojans implies that if Uranus Trojans do exist then they are likely either too small or too dark to be detected.}\par
L$_5$ orbits with $i_0 = 0^{\circ}$, $5^{\circ}$, $15^{\circ}$, and $30^{\circ}$ had half-lives of 103 Myr, 218 Myr, 25 Myr, and 46 Myr, respectively. L$_4$ \textedit{nominal} orbits had a longer half-life than L$_5$ \textedit{nominal} orbits with the same $i_0$. \textedit{This was because nominal orbits in the L$_5$ swarm had significantly larger average initial Trojan resonant angle libration amplitudes than their L$_4$ counterparts with the same $i_0$. The larger libration amplitudes led to more instability.} \par
\newpage
\setcounter{secnumdepth}{0} % Turns off numbering for sections.
\section{Acknowledgements}
The author would like to thank the referee for their invaluable comments and Jonathan Horner for his assistance with this project.
\newpage

\section{Data Availability}
Removal time data is available upon request.

\section{Appendix}
\noindent The function used to find the appropriate bin width of a histogram \textedit{written in MATLAB}:\\
\\
function binWidth = getbindwidthfromraw(rawBinWidth)\\
\\
	powOfTen = 10.\string^(floor(log10(rawBinWidth)));\\
	relSize = rawBinWidth / powOfTen;\\ 
	if relSize $<$ 1.5\par
        binWidth = 1*powOfTen;\\
	elseif relSize $<$ 2.5\par
       binWidth = 2*powOfTen;\\
    elseif relSize $<$ 4\par
        binWidth = 3*powOfTen;\\
    elseif relSize $<$ 7.5\par
        binWidth = 5*powOfTen;\\
    else\par
        binWidth = 10*powOfTen;\\
    end\\
\\
end\\

\noindent The number of bins along with the left and right edges of the histogram are found using:\\
\\
xmin = min(data);\\
xmax = max(data);\\
leftEdge = min(binWidth*floor(xmin ./ binWidth), xmin);\\
$n_b$ = max(1, ceil((xmax-leftEdge) ./ binWidth));\\
rightEdge = max(leftEdge + nbinsActual.*binWidth, xmax);\\


\begin{thebibliography}{}
\bibitem[Alexandersen et al.(2013)]{2013Sci...341..994A} Alexandersen, M., Gladman, B., Greenstreet, S., et al.\ 2013, Science, 341, 994
\bibitem[Barbieri et al.(2003)]{2003MSAIS...3...40B} Barbieri, C., Magrin, S., Marchi, S., et al.\ 2003, Memorie della Societa Astronomica Italiana Supplementi, 3, 40
\bibitem[Borisov et al.(2017)]{2017MNRAS.466..489B} Borisov, G., Christou, A., Bagnulo, S., et al.\ 2017, \mnras, 466, 489
\bibitem[Brouwer(1937)]{1937AJ.....46..149B} Brouwer, D.\ 1937, \aj, 46, 149. doi:10.1086/105423
\bibitem[Chambers(1999)]{ChambersJE:1999} Chambers, J. E. 1999, MNRAS, 304, 793
\bibitem[Connors et al.(2011)]{2011Natur.475..481C} Connors, M., Wiegert, P., \& Veillet, C.\ 2011, \nat, 475, 481
\bibitem[de La Fuente Marcos, \& de La Fuente Marcos(2013)]{2013MNRAS.432L..31D} de La Fuente Marcos, C., \& de La Fuente Marcos, R.\ 2013, \mnras, 432, L31
\bibitem[\protect\citeauthoryear{de la Fuente Marcos \& de la Fuente Marcos}{2014}]{2014MNRAS.439.2970D} de la Fuente Marcos C., de la Fuente Marcos R., 2014, MNRAS, 439, 2970
\bibitem[\protect\citeauthoryear{de la Fuente Marcos \& de la Fuente Marcos}{2017}]{2017MNRAS.467.1561D} de la Fuente Marcos C., de la Fuente Marcos R., 2017, MNRAS, 467, 1561
\bibitem[Di Sisto et al.(2014)]{2014Icar..243..287D} Di Sisto, R.~P., Ramos, X.~S., \& Beaug{\'e}, C.\ 2014, \icarus, 243, 287
\bibitem[Dvorak et al.(2010)]{2010CeMDA.107...51D} Dvorak, R., Bazs{\'o}, {\'A}., \& Zhou, L.-Y.\ 2010, Celestial Mechanics and Dynamical Astronomy, 107, 51. doi:10.1007/s10569-010-9261-y
\bibitem[Fleming, \& Hamilton(2000)]{2000Icar..148..479F} Fleming, H.~J., \& Hamilton, D.~P.\ 2000, \icarus, 148, 479
\bibitem[Gascheau(1843)]{Gascheau...1843}Gascheau, G.\ 1843, C. R. Acad. Sci., 16
\bibitem[{\'E}rdi et al.(2013)]{2013CeMDA.117....3E} {\'E}rdi, B., Forg{\'a}cs-Dajka, E., \& S{\"u}li, {\'A}.\ 2013, Celestial Mechanics and Dynamical Astronomy, 117, 3. doi:10.1007/s10569-013-9492-9
\bibitem[Gerdes et al.(2022)]{2022AAS...24022705G} Gerdes, D., Napier, K., Lin, H.-W., et al.\ 2022, \aas
\bibitem[Gomes(1998)]{1998AJ....116.2590G} Gomes, R.~S.\ 1998, \aj, 116, 2590. doi:10.1086/300582
\bibitem[Gomes \& Nesvorn{\'y}(2016)]{2016A&A...592A.146G} Gomes, R. \& Nesvorn{\'y}, D.\ 2016, \aap, 592, A146. doi:10.1051/0004-6361/201527757
\bibitem[Gradie \& Veverka(1980)]{1980Natur.283..840G} Gradie, J. \& Veverka, J.\ 1980, \nat, 283, 840. doi:10.1038/283840a0
\bibitem[Greenstreet et al.(2013)]{2013DPS....4550802G} Greenstreet, S., Alexandersen, M., Gladman, B., et al.\ 2013, AAS/Division for Planetary Sciences Meeting Abstracts \#45, 508.02
\bibitem[Holman, \& Wisdom(1993)]{1993AJ....105.1987H} Holman, M.~J., \& Wisdom, J.\ 1993, \aj, 105, 1987
\bibitem[Holt et al.(2020)]{2020MNRAS.495.4085H} Holt, T.~R., Nesvorn{\'y}, D., Horner, J., et al.\ 2020, \mnras, 495, 4085. doi:10.1093/mnras/staa1348
\bibitem[Horner \& Evans(2006)]{2006MNRAS.367L..20H} Horner, J. \& Evans, N.~W.\ 2006, \mnras, 367, L20. doi:10.1111/j.1745-3933.2006.00131.x
\bibitem[Horner et al.(2004)]{2004MNRAS.354..798H} Horner, J., Evans, N.~W., \& Bailey, M.~E.\ 2004, \mnras, 354, 798. doi:10.1111/j.1365-2966.2004.08240.x
\bibitem[Horner et al.(2020)]{2020PASP..132j2001H} Horner, J., Kane, S.~R., Marshall, J.~P., et al.\ 2020, \pasp, 132, 102001. doi:10.1088/1538-3873/ab8eb9
\bibitem[Horner, \& Lykawka(2010a)]{2010MNRAS.402...13H} Horner, J., \& Lykawka, P.~S.\ 2010, \mnras, 402, 13
\bibitem[Horner \& Lykawka(2010b)]{2010MNRAS.405...49H} Horner, J. \& Lykawka, P.~S.\ 2010, \mnras, 405, 49. doi:10.1111/j.1365-2966.2010.16441.x
\bibitem[Horner et al.(2012a)]{HornerJ:2012a} Horner, J., Lykawka, P. S., Bannister, M. T., \& Francis, P. 2012a, \mnras, 422, 2145
\bibitem[Horner et al.(2012)]{2012MNRAS.423.2587H} Horner, J., M{\"u}ller, T.~G., \& Lykawka, P.~S.\ 2012, \mnras, 423, 2587
\bibitem[\protect\citeauthoryear{Horner \& Wyn Evans}{2006}]{2006MNRAS.367L..20H} Horner J., Wyn Evans N., 2006, MNRAS, 367, L20
\bibitem[Hou et al.(2016)]{2016CeMDA.125..451H} Hou, X., Scheeres, D.~J., \& Liu, L.\ 2016, Celestial Mechanics and Dynamical Astronomy, 125, 451
\bibitem[Hsieh et al.(2020)]{2020AJ....159..179H} Hsieh, H.~H., Novakovi{\'c}, B., Walsh, K.~J., et al.\ 2020, \aj, 159, 179. doi:10.3847/1538-3881/ab7899
\bibitem[\protect\citeauthoryear{Kortenkamp, Malhotra \& Michtchenko}{2004}]{2004Icar..167..347K} Kortenkamp S.~J., Malhotra R., Michtchenko T., 2004, Icar, 167, 347
\bibitem[Lagrange(1772)]{Lagrange_J_L_1772} Lagrange J.L.\ 1772, Prix de l'acad{\'e}mie royale des Sciences de paris, 9, 292
\bibitem[Laskar(2008)]{2008Icar..196....1L} Laskar, J.\ 2008, \icarus, 196, 1. doi:10.1016/j.icarus.2008.02.017
\bibitem[\protect\citeauthoryear{Levison, Shoemaker \& Shoemaker}{1997}]{1997Natur.385...42L} Levison H.~F., Shoemaker E.~M., Shoemaker C.~S., 1997, Natur, 385, 42
\bibitem[Lin(2018)]{2018ASPC..513...57L} Lin, H.-W.\ 2018, Serendipities in the Solar System and Beyond, 57
\bibitem[\protect\citeauthoryear{Lykawka \& Horner}{2011}]{2011epsc.conf.1246L} Lykawka P.~S., Horner J., 2011, EPSC-DPS Joint Meeting 2011,  1246, epsc.conf
\bibitem[Lykawka et al.(2009)]{2009MNRAS.398.1715L} Lykawka, P.~S., Horner, J., Jones, B.~W., et al.\ 2009, \mnras, 398, 1715
\bibitem[Lykawka et al.(2011)]{2011MNRAS.412..537L} Lykawka, P.~S., Horner, J., Jones, B.~W., et al.\ 2011, \mnras, 412, 537
\bibitem[Ma(2019)]{2019ESS.....431201M} Ma, Y.\ 2019, AAS/Division for Extreme Solar Systems Abstracts 51, 312.01
\bibitem[Marsset et al.(2014)]{2014EPSC....9...54M} Marsset, M., Vernazza, P., Gourgeot, F., et al.\ 2014, European Planetary Science Congress
\bibitem[Mart{\'\i} et al.(2016)]{2016MNRAS.460.1094M} Mart{\'\i}, J.~G., Cincotta, P.~M., \& Beaug{\'e}, C.\ 2016, \mnras, 460, 1094. doi:10.1093/mnras/stw1035
\bibitem[Martin(2022)]{Martin...personal...}Martin, A.\ 2022, \dps
\bibitem[Martin et al.(2017)]{2017AGUFM.P23A2708M} Martin, A., Emery, J.~P., \& Lindsay, S.~S.\ 2017, AGU Fall Meeting Abstracts
\bibitem[\protect\citeauthoryear{Marzari, Farinella, Davis, Scholl \& Campo Bagatin}{1997}]{1997Icar..125...39M} Marzari F., Farinella P., Davis D.~R., Scholl H., Campo Bagatin A., 1997, Icar, 125, 39
\bibitem[\protect\citeauthoryear{Marzari \& Scholl}{2002}]{2002Icar..159..328M} Marzari F., Scholl H., 2002, Icar, 159, 328
\bibitem[Marzari \& Scholl(2007)]{2007MNRAS.380..479M} Marzari, F. \& Scholl, H.\ 2007, \mnras, 380, 479. doi:10.1111/j.1365-2966.2007.12095.x
\bibitem[Marzari et al.(1996)]{1996Icar..119..192M} Marzari, F., Scholl, H., \& Farinella, P.\ 1996, \icarus, 119, 192
\bibitem[Marzari et al.(2002)]{2002aste.book..725M} Marzari, F., Scholl, H., Murray, C., et al.\ 2002, Asteroids III, 725
\bibitem[\protect\citeauthoryear{Marzari, Tricarico \& Scholl}{2002}]{2002ApJ...579..905M} Marzari F., Tricarico P., Scholl H., 2002, ApJ, 579, 905
\bibitem[Marzari et al.(2002)]{2002aste.book..725M} Marzari, F., Scholl, H., Murray, C., et al.\ 2002, Asteroids III, 725
\bibitem[Marzari et al.(2003)]{2003A&A...410..725M} Marzari, F., Tricarico, P., \& Scholl, H.\ 2003, \aap, 410, 725
\bibitem[\protect\citeauthoryear{Marzari, Tricarico \& Scholl}{2003}]{2003Icar..162..453M} Marzari F., Tricarico P., Scholl H., 2003, Icar, 162, 453
\bibitem[Marzari et al.(2003)]{2003MNRAS.345.1091M} Marzari, F., Tricarico, P., \& Scholl, H.\ 2003, \mnras, 345, 1091. doi:10.1046/j.1365-2966.2003.07051.x
\bibitem[MATLAB(2018)]{2018MATLAB...R2018a}MATLAB 2018, MATLAB version 9.4.0.949201 R2018a (The Mathworks, Inc. Natick, Massachusetts)
\bibitem[Morbidelli et al.(2005)]{2005Natur.435..462M} Morbidelli, A., Levison, H.~F., Tsiganis, K., et al.\ 2005, \nat, 435, 462. doi:10.1038/nature03540
\bibitem[Murray, \& Dermott(1999)]{1999ssd..book.....M} Murray, C.~D., \& Dermott, S.~F.\ 1999, Solar system dynamics by C.D. Murray and S.F. McDermott. (Cambridge)
\bibitem[Namouni(1999)]{1999Icar..137..293N} Namouni, F.\ 1999, \icarus, 137, 293. doi:10.1006/icar.1998.6032
\bibitem[Nesvorny(2000)]{2000DPS....32.1902N} Nesvorny, D.\ 2000, \dps
\bibitem[\protect\citeauthoryear{Nesvorn{\'y} \& Dones}{2002}]{2002Icar..160..271N} Nesvorn{\'y} D., Dones L., 2002, Icar, 160, 271
\bibitem[Nesvorn{\'y} \& Vokrouhlick{\'y}(2009)]{2009AJ....137.5003N} Nesvorn{\'y}, D. \& Vokrouhlick{\'y}, D.\ 2009, \aj, 137, 5003. doi:10.1088/0004-6256/137/6/5003
\bibitem[Nesvorn{\'y} et al.(2018)]{2018NatAs...2..878N} Nesvorn{\'y}, D., Vokrouhlick{\'y}, D., Bottke, W.~F., et al.\ 2018, Nature Astronomy, 2, 878. doi:10.1038/s41550-018-0564-3
\bibitem[Park et al.(2021)]{2021AJ....161..105P} Park, R.~S., Folkner, W.~M., Williams, J.~G., et al.\ 2021, \aj, 161, 105. doi:10.3847/1538-3881/abd414
\bibitem[Parker(2015)]{2015Icar..247..112P} Parker, A.~H.\ 2015, \icarus, 247, 112. doi:10.1016/j.icarus.2014.09.043
\bibitem[Polishook et al.(2017)]{2017NatAs...1E.179P} Polishook, D., Jacobson, S.~A., Morbidelli, A., et al.\ 2017, Nature Astronomy, 1, 0179
\bibitem[Pirani et al.(2019)]{2019A&A...631A..89P} Pirani, S., Johansen, A., \& Mustill, A.~J.\ 2019, \aap, 631, A89
\bibitem[Rein \& Liu(2012)]{2012A&A...537A.128R} Rein, H., \& Liu, S.-F.\ 2012, A\&A, 537, A128 
\bibitem[Rein \& Spiegel(2015)]{2015MNRAS.446.1424R} Rein, H., \& Spiegel, D.~S.\ 2015, MNRAS, 446, 1424 
\bibitem[Scholl et al.(2005)]{2005Icar..175..397S} Scholl, H., Marzari, F., \& Tricarico, P.\ 2005, \icarus, 175, 397
\bibitem[Scott(2010)]{Scott...2010Wiley}Scott D., 2010, Wiley Interdisciplinary Reviews: Computational Statistics. 2. 10.1002/wics.103.
\bibitem[Sheppard \& Trujillo(2006)]{2006DPS....38.4403S} Sheppard, S.~S. \& Trujillo, C.\ 2006, \dps
\bibitem[Sicardy(2010)]{2010CeMDA.107..145S} Sicardy, B.\ 2010, Celestial Mechanics and Dynamical Astronomy, 107, 145. doi:10.1007/s10569-010-9259-5
\bibitem[Slyusarev et al.(2020)]{2020IAUS..345..345S} Slyusarev, I., Glezina, D., \& Belskaya, I.\ 2020, Origins: From the Protosun to the First Steps of Life, 345, 345. doi:10.1017/S1743921319001959
\bibitem[Souami \& Souchay(2012)]{2012A&A...543A.133S} Souami, D. \& Souchay, J.\ 2012, \aap, 543, A133. doi:10.1051/0004-6361/201219011
\bibitem[Trilling et al.(2021)]{2021DPS....5320204T} Trilling, D., Gerdes, D., Fuentes, C., et al.\ 2021, \dps
\bibitem[Tsiganis et al.(2005a)]{2005Natur.435..459T} Tsiganis, K., Gomes, R., Morbidelli, A., et al.\ 2005, \nat, 435, 459
\bibitem[Tsiganis et al.(2005)]{2005CeMDA..92...71T} Tsiganis, K., Varvoglis, H., \& Dvorak, R.\ 2005, Celestial Mechanics and Dynamical Astronomy, 92, 71. doi:10.1007/s10569-004-3975-7
\bibitem[Volk et al.(2016)]{2016AJ....152...23V} Volk, K., Murray-Clay, R., Gladman, B., et al.\ 2016, \aj, 152, 23. doi:10.3847/0004-6256/152/1/23
\bibitem[Wisdom(1983a)]{1983Icar...56...51W} Wisdom, J.\ 1983, \icarus, 56, 51
\bibitem[Wisdom(1983b)]{1983Metic..18..422W} Wisdom, J.\ 1983, Meteoritics, 18, 422
\bibitem[\protect\citeauthoryear{Wolf}{1906}]{1906AN....170..353W} Wolf M., 1906, AN, 170, 353
\bibitem[Wood(2019)]{2019dsss.book.....W} Wood, J.\ 2019, The Dynamics of Small Solar System Bodies
\bibitem[Wood et al.(2017)]{2017AJ....153..245W} Wood, J., Horner, J., Hinse, T.~C., et al.\ 2017, \aj, 153, 245. doi:10.3847/1538-3881/aa6981
\bibitem[Wood et al.(2018)]{2018AJ....155....2W} Wood, J., Horner, J., Hinse, T.~C., et al.\ 2018, \aj, 155, 2. doi:10.3847/1538-3881/aa9930
\bibitem[Yoshida et al.(2019)]{2019P&SS..169...78Y} Yoshida, F., Terai, T., Ito, T., et al.\ 2019, \planss, 169, 78
\bibitem[Zhou et al.(2020)]{2020A&A...633A.153Z} Zhou, L., Zhou, L.-Y., Dvorak, R., et al.\ 2020, \aap, 633, A153. doi:10.1051/0004-6361/201936332
\end{thebibliography}
\end{document}